\documentclass[twocolumn,aps,prd,showpacs,nofootinbib,eqsecnum]{revtex4}

\usepackage{amsmath, amsfonts, amssymb, amsxtra}

\newcommand{\ve}[1]{{\bf #1}}
\newcommand{\Long}{\mathbb{L}}
\newcommand{\barLong}{\bar{\Long}}
\newcommand{\dnachd}[2]{\frac{\partial #1}{\partial #2}}
\newcommand{\Lie}{\mathcal{L}}

\newcommand{\gfour}{{\mbox{\bf g}}}
\newcommand{\M}{\mathcal{M}}
\newcommand{\V}{\mathcal{V}} 
\newcommand{\CF}{\varphi}  
\newcommand{\trK}{\tau}    
\newcommand{\pperp}{{\perp\!\!\!\:\!\!\perp}}

\begin{document}

\title{Extrinsic Curvature and the Einstein Constraints}

\author{Harald P. Pfeiffer}
\author{James W. York, Jr.}
\affiliation{Department of Physics, Cornell University, Ithaca, New York\ \ 14853}

\date{\today}

\begin{abstract}
The Einstein initial-value equations in the extrinsic curvature
(Hamiltonian) representation and conformal thin sandwich (Lagrangian)
representation are brought into complete conformity by the use of a
decomposition of symmetric tensors which involves a weight function.
In stationary spacetimes, there is a natural choice of the weight
function such that the transverse traceless part of the extrinsic
curvature (or canonical momentum) vanishes.
\end{abstract}
\maketitle

\section{Introduction}

In this paper we introduce a new decomposition of symmetric tensors
and apply it to the construction of extrinsic curvature in the
initial-value equations of general relativity.  Our results improve
previous work that dealt with extrinsic curvature \cite{York:1979,
Murchadha-York:1974a}.  The new findings are consistent with the
conformal thin sandwich equations, which involve no tensor
decompositions \cite{York:1999}. 

The thin sandwich and extrinsic curvature formulations differ in
whether the velocity or the momentum associated with the conformal
spatial metric is specified.  The corresponding Lagrangian and
Hamiltonian pictures of dynamics must certainly agree, and we find a
clear and explicit form of such agreement in our analysis.  This would
not occur without the presence of a weight function, which must be the
lapse function, in the new decomposition.

The presence of the lapse function is also crucial for another result
of this decomposition: For stationary spacetimes, there is a natural
way to choose the lapse function which results in a {\em vanishing}
transverse traceless part of the extrinsic curvature.

Finally, we extend the conformal thin sandwich equations by giving the
velocity $\dot\trK$ of the mean curvature $\trK=\mbox{Tr}\,\ve{K}$ in
order to determine the lapse function.

\section{New Tensor Splittings}

We will define and apply a new class of covariant decompositions of
the extrinsic curvature\footnote{We give a general definition of
extrinsic curvature in the appendix, which also details our
conventions.  See in particular Eqns.~(\ref{eq:metric}) and
(\ref{eq:dtgij}).}  of a three-dimensional hypersurface $\M$.  (These
decompositions were introduced informally by the second author in
March, 2001.)  It should be noted that they apply to {\em any}
symmetric $\binom{2}{0}$ or $\binom{0}{2}$ tensor for {\em any} dimension
$m\ge 3$.  Generalization to $m>3$ is straightforward (the case $m=2$
is special).  The decompositions use the geometry of $\M$, that is,
the metric $\bar g_{ij}$ and the derivative $\bar\nabla$, together
with a positive scalar function $\bar\sigma$ to be specified later.

Given $(\M, \bar g)$, we first remove the trace of the
extrinsic curvature $\bar K^{ij}$:
\begin{equation}\label{eq:K-traceless}
  \bar K^{ij}=\bar A^{ij}+\frac{1}{3}\trK\bar g^{ij}
\end{equation}
where $\bar A^{ij}$ is traceless and $\trK=\bar\trK=\bar g_{ij}\bar
K^{ij}$ is the trace of $\bar K^{ij}$, called the ``mean curvature.''
Note that here, and in the sequel, over-bars are used to distinguish
spatial tensors from their conformally transformed counterparts.  For
example, $\bar g_{ij}=\CF^4g_{ij}$, $\CF>0$, where $\bar g$ and $g$
are both spatial metrics.  Quantities with over-bars have physical
values.

The next step is to decompose the traceless symmetric tensor $\bar
A^{ij}$ covariantly.  Like previous studies, for example
\cite{Deser:1967, York:1973, York:1974}, we attempt to find a
covariantly defined divergence-free part of $\bar A^{ij}$ with zero
trace.  We want to stress that this so-called ``transverse-traceless''
(TT) part of $\bar K^{ij}$ (or $\bar A^{ij}$) is {\em not unique}.
There are infinitely many mathematically well-defined ways to extract
such a piece of the original tensor, for example by varying the choice
of a weight function (see (\ref{eq:TT}) below).  However, it is
possible that by imposing {\em geometrical or physical requirements},
we can make the TT part and the other parts unique.

We write $\bar A^{ij}$ as a sum of a TT part and a weighted longitudinal
(vector) traceless part as follows:
\begin{equation}\label{eq:TT}
\bar A^{ij}=\bar A^{ij}_{TT}+\bar\sigma^{-1}(\barLong Y)^{ij}
\end{equation}
where the inverse weight function $\bar\sigma$ is a uniformly positive
and bounded scalar on $\M$: $0<\epsilon\le\bar\sigma<\infty$, 
$\epsilon=\mbox{constant.}$ We define \cite{Deser:1967, York:1973, York:1974}
\begin{equation}\label{eq:L}
  (\barLong Y)^{ij}
  =\bar\nabla^iY^j+\bar\nabla^jY^i-\frac{2}{3}\bar g^{ij}\bar\nabla_kY^k,
\end{equation}
which is proportional to the Lie derivative with respect to $Y^i$ of a
unimodular inverse metric conformally related to $\bar g^{ij}$; thus,
take the Lie derivative of $\left[\det(\bar g_{kl})\right]^{1/3}\bar
g^{ij}$.  Expression (\ref{eq:L}) is zero for $Y^i\neq 0$ if and only
if $Y^i$ is a conformal Killing vector of the metric, if such a
symmetry exists.  This also suggests that if we make a conformal
transformation $\bar g_{ij}\to\CF^{-4}\bar g_{ij}=g_{ij}$, $\CF>0$,
then $Y^i$ will not be conformally scaled: $Y^i\to Y^i$.  We will
adopt this scaling rule below; because of its simplicity, we omit
over-bars on vectors like $Y^i$.

The decomposition is effected by solving for $Y^i$ in 
\begin{equation}\label{eq:DivL}
\bar\nabla_j\left[\;\bar\sigma^{-1}\,(\barLong Y)^{ij}\;\right]
=\bar\nabla_j\bar A^{ij}
\end{equation}
and then setting
\begin{equation}
  \bar A^{ij}_{TT}\equiv
\bar A^{ij}
-\bar\sigma^{-1}\left(\barLong Y_{
                      \mbox{\footnotesize (solution)}}\right)^{ij},
\end{equation}
where the given $\bar\sigma$ and the solution for $Y^i$ are inserted.

The operator on the left side of (\ref{eq:DivL}),
$(\bar\Delta_{L,\bar\sigma}Y)^i$, is similar to the vector Laplacian
$(\bar\Delta_LX)^i\equiv\bar\nabla_j(\barLong Y)^{ij}$, which is
solvable on compact manifolds and on asymptotically flat manifolds,
given certain asymptotic conditions.  The conditions needed are not
onerous.  Inserting the weight factor leaves the operator in
``divergence form,'' and does not affect the formal self-adjointness
of the weighted vector Laplacian.  The calculation of self-adjointness
is carried out in the natural measure $\mu_{\bar g}=\sqrt{\bar
g\,}\,d^3x$.  On the other hand, the two pieces of $\bar A^{ij}$ in
(\ref{eq:TT}) are formally orthogonal in the positive measure
$\bar\mu_{\bar g,\bar\sigma}=\bar\sigma\sqrt{\bar
g\,}\,d^3x=\bar\sigma\,\mu_{\bar g}$.

Note that (\ref{eq:DivL}) must be supplemented with {\em boundary
conditions} in the case where $\M$ is not compact without boundary and
that the solution of (\ref{eq:DivL}) and $\bar A^{ij}_{TT}$ will
depend on these boundary
conditions\cite{Pfeiffer-Cook-Teukolsky:2002}.  This caveat about
boundary conditions is very serious in practice, where there may be
excised regions of $\M$, and where $\M$ may have no asymptotic region,
though it models a space with Euclidean asymptotic conditions. There
is no uniqueness without boundary conditions!

The conformal properties of the new splittings assigned here are
very important for application to the Einstein constraints, but they are
also very interesting in themselves.

We take 
\begin{gather}
\label{eq:ConfScalings1a}
\bar A^{ij}_{TT}=\CF^{-10}A^{ij}_{TT},\\
\label{eq:ConfScalings1b}
\bar Y^i=Y^i,\\
\label{eq:ConfScalings1c}
\bar\sigma=\CF^6\sigma.
\end{gather}
Transformations (\ref{eq:ConfScalings1a}) and
(\ref{eq:ConfScalings1b}) should be familiar.  We adopted
(\ref{eq:ConfScalings1c}) to obtain correct divergence relations in
the sequel.  It also maintains the correct transformations when we set
$\bar\sigma=2\bar N$ and $\sigma=2N$ in Section \ref{sec:IV} below.

Next we use $\bar g_{ij}=\CF^4g_{ij}$ with its concomitant
transformation rule
\begin{equation}\label{eq:ConfScalings2}
\bar\Gamma^i_{jk}
=\Gamma^i_{jk}+2\CF^{-1}\left(\delta^i_j\partial_k\CF+\delta^i_k\partial_j\CF
                              -g_{jk}g^{il}\partial_l\CF\right)
\end{equation}
for the Christoffel symbols of $\bar g_{ij}$ and $g_{ij}$.
If we were to consider a general symmetric tensor $\bar T^{ij}$
in three dimensions, with $\bar g_{ij}=\CF^4g_{ij}$, then $\bar
T^{ij}=\CF^xT^{ij}$ yields
\begin{equation}\label{eq:general-transform}
\begin{aligned}
\bar\nabla_j\bar T^{ij}
  =\CF^x\Big[ &(\nabla_jT^{ij})+(x+10)(\partial_j\log\CF)T^{ij}\\
   &-2(\partial_j\log\CF)\left(g^{ij}g_{kl}T^{kl}\right)\Big].
\end{aligned}
\end{equation}
This shows why we choose $x=-10$ in (\ref{eq:ConfScalings1a}), as well
as the zero trace, $g_{ij}\bar A^{ij}_{TT}=0$.

From (\ref{eq:ConfScalings1a})--(\ref{eq:general-transform}), we find,
besides $\bar\nabla_j\bar A^{ij}_{TT}=\CF^{-10}\nabla_jA^{ij}_{TT}=0$,

\begin{align}\label{eq:ConfScalings3a}
  (\barLong Y)^{ij}&=\CF^{-4}(\Long Y)^{ij},\\
\bar\sigma^{-1}(\barLong Y)^{ij}
&=\CF^{-10}\left[\,\sigma^{-1}(\Long Y)^{ij}\right].
\end{align}
Therefore, 
\begin{align}
\bar A^{ij}&=\bar A^{ij}_{TT}+\bar\sigma^{-1}(\barLong Y)^{ij}            
\nonumber\\
\label{eq:Aij-transformation}
  &=\CF^{-10}\left[A^{ij}_{TT}+\sigma^{-1}(\Long Y)^{ij}\right]
=\CF^{-10}A^{ij}.
\end{align}
Thus, 
\begin{equation}
\bar K^{ij}
=\CF^{-10}\left[A^{ij}_{TT}+\sigma^{-1}(\Long Y)^{ij}\right]
  +\frac{1}{3}\CF^{-4}g^{ij}\trK.
 \end{equation}
We note that the conformal quantities $A^{ij}_{TT}$ and
$\sigma^{-1}(\Long Y)^{ij}$ are orthogonal in the re-scaled measure
$\sigma\sqrt{g\,}d^3x=\sigma\mu_g=\mu_{g,\sigma}$.

\section{Einstein Constraints}

The extrinsic curvature $\bar K^{ij}$ is uniquely related to the
canonical momentum
\begin{equation}
\bar\pi^{ij}
=(\mbox{const.})\bar g^{1/2}\left(\trK \bar g^{ij}-\bar K^{ij}\right)
\end{equation}
Therefore the present study is devoted to a construction of quantities
belonging to the canonical or Hamiltonian picture of dynamics, while
the conformal thin sandwich equations\cite{York:1999} belong to the
velocity phase space or Lagrangian picture.  It is essential to
understand the constraints in {\em both} pictures, and the pictures
must be geometrically and physically consistent.

Elaboration of the constraints in vacuum
\begin{align}
\bar\nabla_j\left(\trK\bar g^{ij}-\bar K^{ij}\right)&=0,\\ \bar
K_{ij}\bar K^{ij}-\trK^2-\bar R&=0,\label{eq:scalar-constraint0}
\end{align}
where $\bar R$ is the ``scalar curvature'' or ``trace of the Ricci
tensor'' of $(\M,\bar g,\bar\nabla)$, is facilitated by displaying them as
\begin{gather}
\label{eq:vector-constraint}
\bar\nabla_j\bar A^{ij}=\frac{2}{3}\bar g^{ij}\partial_j\trK,\\
\label{eq:scalar-constraint}
\bar A_{ij}\bar A^{ij}-\frac{2}{3}\trK^2-\bar R=0.
\end{gather}
The form (\ref{eq:vector-constraint}) and (\ref{eq:scalar-constraint})
of the constraints is the better one for beginning formulating the
constraints in the thin sandwich decomposition or in the extrinsic
curvature form.

It has been pointed out \cite{Murchadha:1973, York:1979} that the
standard transverse traceless tensor decomposition \cite{York:1973,
York:1974},
\begin{equation}\label{eq:old-TT-decomp}
\bar T^{ij}-\frac{1}{3}\bar g^{ij}\left(\bar g_{kl}\bar T^{kl}\right)
= \bar T^{ij}_{TT}+(\barLong V)^{ij}
\end{equation}
has the property that extracting the TT part of a symmetric tensor
does not commute with conformal transformations: Specifically,
(\ref{eq:old-TT-decomp}), which is simply (\ref{eq:K-traceless}) and
(\ref{eq:TT}) with the weight $\bar\sigma^{-1}$ entirely ignored,
produces parts which transform as (\ref{eq:ConfScalings1a}) and
(\ref{eq:ConfScalings3a}) under conformal transformations.  The two
parts, therefore, do not transform alike.

The fact that (\ref{eq:ConfScalings1a}) and (\ref{eq:ConfScalings3a})
behave differently under conformal transformations leads to two
inequivalent methods of decomposing $\bar A^{ij}$.  What
Isenberg\cite{Isenberg:1974} has called ``Method A'' is first to
transform $\bar A^{ij}$ conformally, then split it with $\sigma$ and
$\bar\sigma$ ignored, then transform the unbarred vector part with the
``wrong'' transformation (using $\CF^{-10}$ instead of $\CF^{-4}$ so
as to scale its divergence).  This is discussed in \cite{York:1979}.
``Method B'' is, in effect, to split $\bar A^{ij}$ first, then
transform conformally to obtain a slightly different, and more
difficult, form of the constraints \cite{Murchadha-York:1974a}.  The
existence of {\em two} methods of dealing with the extrinsic curvature
using the old tensor splitting suggests that neither is the optimal
method.  The method introduced here has no such ambiguity and can be
regarded as the resolution (in the conformal framework) of the initial
value problem in the extrinsic curvature representation.

We apply the splitting (\ref{eq:TT}) to the vector constraint
(\ref{eq:vector-constraint}).  After the conformal transformations, 
we obtain
\begin{equation}\label{eq:vector-constraint2}
\nabla_j\left[\sigma^{-1}(\Long B)^{ij}\right]=\frac{2}{3}\CF^6\nabla^i\trK,
\end{equation}
where we used
\begin{equation}
A^{ij}=A^{ij}_{TT}+\sigma^{-1}(\Long B)^{ij}.
\end{equation}
Equation (\ref{eq:vector-constraint2}) is the momentum constraint.  We
will see that it will determine $B^i$ once $\sigma$ is chosen.  The
term $A^{ij}_{TT}$ disappears from the vector constraint.  It is
``freely specifiable'' and can be determined by extracting the TT part
of some traceless symmetric tensor $C^{ij}$:
\begin{equation}\label{eq:A^ij_TT}
A^{ij}_{TT}=C^{ij}-\sigma^{-1}(\Long V)^{ij},
\end{equation}
where $C^{ij}$ is freely given, as is $\sigma>0$, and $V^i$ is then
determined by solving an equation similar to (\ref{eq:DivL}).  In this
step boundary conditions must be applied when $\M$ has boundaries,
which will influence $V^i$ and $A^{ij}_{TT}$.

Thus $C^{ij}$ supplies a ``source'' in the vector constraint
(\ref{eq:vector-constraint}), which, if we define \cite{Cantor:1980,
Cook:2000}
\begin{equation}
X^i=B^i-V^i,
\end{equation}
becomes (in vacuum)
\begin{equation}\label{eq:vector-constraint3}
\nabla_j\left[\sigma^{-1}(\Long X)^{ij}\right]
=\frac{2}{3}\CF^6\nabla^i\trK-\nabla_{\!j}C^{ij}.
\end{equation}
From its solution we can construct
\begin{align}\label{eq:Aij}
A^{ij}=& \;C^{ij}+\sigma^{-1}(\Long X)^{ij}\\
=& \;A^{ij}_{TT}+\sigma^{-1}(\Long B)^{ij}.
\end{align}

Next, recall that $\bar g_{ij}=\CF^4g_{ij}$
implies\cite{Lichnerowicz:1944}
\begin{equation}\label{eq:conformal-R}
\bar R=R\CF^{-4}-8\CF^{-5}\Delta\CF,
\end{equation}
where $R$ is the scalar curvature of $g_{ij}$.
Eq.~(\ref{eq:conformal-R}) enables us to rewrite the scalar constraint
(\ref{eq:scalar-constraint}) in transformed variables as
\begin{equation}\label{eq:scalar-constraint3}
\Delta\CF-\frac{1}{8}R\CF
=-\frac{1}{8}A_{ij}A^{ij}\CF^{-7}+\frac{1}{12}\trK^2\CF^5.
\end{equation}
(We also used Eq. (\ref{eq:Aij-transformation}), which implies 
$\bar A_{ij}=\CF^{-2}A_{ij}$.)

\section{The Weight Function and the Lapse Function}
\label{sec:IV}

The extrinsic curvature formulation of the initial value problem uses,
in essence, the canonical variables.  Furthermore, it depends only on
the embedding (encoded by $\bar K_{ij}$) of the hypersurface $(\M,
\bar g)$ into the four-dimensional spacetime $(\V, \gfour)$.  It does
not depend on the foliation or the time vector $\partial/\partial t$,
that is, lapse $\bar N$ and shift $\beta^i$.  (This is well
known but can still be a source of confusion.  Therefore, we review
the second fundamental form and the extrinsic curvature in the
Appendix.)  Such a statement cannot be made in the thin sandwich
formulation, where one is interested precisely in the extension along
curves tangent to $\partial/\partial t$.

Since the establishment of the canonical form of the action by
Arnowitt, Deser, and Misner (ADM) \cite{Arnowitt-Deser-Misner:1962},
the shift $\beta^i$ has been taken to be the undetermined multiplier
of the momentum (vector) constraint, while the lapse has been taken to
be the undetermined multiplier of the Hamiltonian (scalar) constraint.
However, it has come to light that the lapse as a multiplier must be
replaced by the {\em lapse antidensity} $\alpha$, a scalar of weight
$(-1)\;$
\cite{Choquet-Bruhat-Ruggeri:1983,Teitelboim:1983,Ashtekar:1987,Anderson-York:1998}.
This replacement is required in order that the canonical framework as
a whole for Einsteinian gravity makes complete sense, that is, that it
works in the same way as for other physical systems that can be
derived from an action principle. For technical details, see
\cite{Anderson-York:1998, York:1999b}.  We are requiring essentially
just that the ``Hamiltonian vector field'' be defined without
reference to the constraints in the whole phase space of $g$'s and
$\pi$'s.

That $\alpha$ and $\beta^i$ are undetermined multipliers of the vector
and scalar constraints means that they are both conformally invariant:
$\bar\alpha=\alpha$ and $\bar\beta^i=\beta^i$.  But the invariance of
$\alpha$ has very interesting consequences.  {\em When the scalar
constraint is satisfied}, then upon examination of the ADM action, we
see that $\alpha=\bar N\bar g^{-1/2}$.  Hence, $\bar\alpha=\alpha$
implies that
\begin{equation}\label{eq:N-scaling}
  \bar N=\CF^6N
\end{equation}
because $\bar g^{1/2}=\CF^6 g^{1/2}$.  Thus, the physical lapse is not
quite arbitrary for our purpose of solving the constraints.  We have a
``trial'' lapse $N$ and a final physical lapse $\bar N$ that will be
determined by $N$ and the solution $\CF$ of the scalar constraint. 

Recall, that in studies of hyperbolic forms of the Einstein
evolution equations with {\em physical} characteristic speeds, it is
also found that $\alpha$, not $\bar N$ is arbitrary (see, {\em e.g.},
\cite{Choquet-Bruhat-York:1997, Kidder-Scheel-Teukolsky:2001}).

From (\ref{eq:N-scaling}) we observe that $\bar N$ and $N$ are related
just as were $\bar\sigma$ and $\sigma$.  Thus we have a natural
geometrical choice:
\begin{equation}\label{eq:sigma=2N}
  \bar\sigma=2\bar N;\qquad\sigma=2N.
\end{equation}
The factors ``$2$'' are chosen for later convenience.  As consequence
of (\ref{eq:sigma=2N}), $\bar N$ appears in $\bar A^{ij}$ and $\bar
K^{ij}$.  Before, however, we have noted that $\bar K^{ij}$ is
independent of $\bar N$.  But our use of $\bar N$ only determines the
{\em splitting} of $\bar K^{ij}$, not $\bar K^{ij}$ itself.

Let us examine the consequences of the choices (\ref{eq:sigma=2N}).
Constraints (\ref{eq:vector-constraint3}) and
(\ref{eq:scalar-constraint3}) become
\begin{gather}
\label{eq:vector-constraint4}
\nabla_j\left[(2N)^{-1}(\Long X)^{ij}\right]
=\frac{2}{3}\CF^6\nabla^i\trK-\nabla_j C^{ij},\\
\label{eq:scalar-constraint4}
\Delta\CF-\frac{1}{8}R\CF=\frac{1}{12}\trK^2\CF^5-\frac{1}{8}A_{ij}A^{ij}\CF^{-7},
\end{gather}
where $A_{ij}$ is given by (\ref{eq:Aij}) with $(2N)$ replacing
$\sigma$.

Of course, $N$ itself is yet to be chosen.  But we have several
remarkable automatic consequences of solving
(\ref{eq:vector-constraint4}) and (\ref{eq:scalar-constraint4}) for
$X^i$ and $\CF$, {\em given} a uniformly positive $N$ and supposing
$\CF$ is also uniformly positive.

\begin{enumerate}
\item We have from (\ref{eq:sigma=2N}) and the conformal transformation
rules, that
\begin{equation}
A^{ij}=A^{ij}_{TT}+(2N)^{-1}(\Long B)^{ij}
\end{equation}
implies
\begin{equation}\label{eq:Aij4}
\bar A^{ij}=\bar A^{ij}_{TT}+(2\bar N)^{-1}(\barLong B)^{ij}.
\end{equation}
The consequence is that $\bar A^{ij}_{TT}$ and $(\barLong B)^{ij}$ are
now orthogonal in the measure $\mu_{\bar g, \bar N}=\bar N\bar
g^{1/2}\,d^3x=\sqrt{-\gfour\,}\,d^3x$, where $\gfour$ is the
determinant of the physical {\em spacetime} metric.  This is the
spacetime measure (apart from ``$dt$''), and so {\em orthogonality} is
determined by the spacetime geometry of $\M$, including the lapse
function, embedded in $\V$.  This result is independent of the method
of determination of $N$.

\item Whatever the result for $\CF$, $X^i$, and $\bar N=\CF^6 N$, we
can show consistency of the results with the conformal thin sandwich
equations \cite{York:1999}.  Consider the traceless inverse metric
velocity $\bar u^{ij}=\partial_t\bar g^{ij}-\frac{1}{3}\bar g^{ij}\bar
g_{kl} \partial_t\bar g^{kl}$, which is related to the traceless
metric velocity $\bar u_{ij}=\partial_t\bar g_{ij}-\frac{1}{3}\bar
g_{ij}\bar g^{kl}\partial_t\bar g_{kl}$ by
\begin{equation}
  \bar u^{ij}=-\bar g^{il}\bar g^{jk}\bar u_{kl}.  
\end{equation}
Using
the evolution equation (\ref{eq:dtgij}), as well as (\ref{eq:Aij4}), we find
\begin{equation}
\bar u^{ij}
=2\bar N\!\left[\bar A^{ij}_{TT}+(2\bar N)^{-1}\left(\barLong B\right)^{ij}
\right]\!-\!(\barLong\beta)^{ij},
\end{equation}
where $\beta^i$ is the shift, which, like the lapse antidensity
$\alpha=Ng^{-1/2}$ can be freely chosen.  The shift {\em vector} is
conformally invariant.  Therefore,
\begin{align}
\bar u^{ij}
&=(2N\CF^6)\left(\CF^{-10} A^{ij}_{TT}\right)
  +\CF^{-4}\left[\Long(B-\beta)\right]^{ij}\\
&=\CF^{-4}\left(2N A^{ij}_{TT}+\left[\Long(B-\beta)\right]^{ij}\right)\\
&=\CF^{-4}\left(2N C^{ij}+\left[\Long(B-V-\beta)\right]^{ij}\right).
\end{align}
We see that three vectors enter $\bar u^{ij}$: $\beta^i$, a gauge
choice, $V^i$, which removes the longitudinal piece of $C^{ij}$, and
$B^i$, which solves the vector constraint.  Let us set
$B^j-V^j-\beta^j=Z^j$.  By the choice of shift $\beta^j=B^j-V^j$, we
could render $Z^j=0$. Then $\bar u^{ij}=\CF^{-4}\left(2 N
C^{ij}\right)$ and $\bar u_{ij}=-\CF^4(2NC_{ij})$.  Furthermore, with
the choice $\beta^i=B^j-V^j=X^j$, the constraints
(\ref{eq:vector-constraint3}) and (\ref{eq:scalar-constraint3}) are
{\em identical} to the constraint equations in the conformal thin
sandwich formalism (Eqs. (14) and (15) of \cite{York:1999}), provided
one identifies $-2NC^{ij}$ with the conformal metric velocity
$u^{ij}$.

\item With $\alpha(x,t)$ given, and $\bar N$ determined by
$\alpha\equiv\bar g^{-1/2}\bar N$, we find that $\bar N$ always obeys
a generalized harmonic evolution,
\begin{equation}\label{eq:harmonicN}
\hat\partial_0\bar N+\bar N^2\trK=\bar N\hat\partial_0\log\alpha,
\end{equation}
where $\hat\partial_0=\partial_t-\Lie_\beta$.
 \end{enumerate}

The identification $\sigma=2N$ still leaves us with the question of
how to choose a trial lapse function in a reasonable way.  In
principle, any choice will do and we will know the geometric meaning
of $\bar N=\CF^6 N$ .

The choice $N=1$, for example, is allowed in the extrinsic curvature
representation.  Indeed, $N=1$ would then give back method A, but now
that the behavior of the lapse function is understood, it gives a
correct $\partial_tg_{ij}$. It also tells us $\bar N=\CF^6=(\bar
g/g)^{1/2}$.  Another choice of $N$ is discussed in Section
\ref{sec:VI} below.

In picking the ``true degrees of freedom,'' that is, the conformal
class of the metric and $\bar K^{TT}_{ij}$, and the mean curvature
$\trK$, the lapse intervenes in pinpointing $\bar K^{TT}_{ij}$.  There
are infinitely many choices other than $N=1$.  This feature did not
arise in previous studies of the extrinsic curvature picture; but, on
the other hand, those pictures do not in general fit the purely
geometric construction of the conformal thin sandwich equations.  In
curved spacetime, the dynamical degrees of freedom in the Hamiltonian
picture do not allow themselves to be identified without the foliation
being inextricably involved.  Another consequence is that time is not
found among the traditional canonical variables alone, even in general
relativity.

\section{Stationary Spacetimes}

Consider a stationary solution of Einstein's equations with timelike
Killing vector $t$.  Given a spacelike hypersurface $\Sigma$, there is
a preferred gauge such that the time-vector of an evolution on
$\Sigma$ coincides with $t$, namely $\bar N=-\!<\!n, t\!>_\gfour$,
$\beta=\pperp t$, where $n$ is the unit normal to $\Sigma$, and
$\pperp$ is the projection operator into $\Sigma$,
$<\!\!\pperp,n\!>_\gfour=0$.  With this choice of lapse and shift, $\bar
g_{ij}$ and $\bar K_{ij}$ will be independent of time.  Using
$\partial_t\bar g_{ij}=0$ in (\ref{eq:Kij}) yields
\begin{equation}
\bar K_{ij}
=\frac{1}{2\bar N}\left(\bar\nabla_i\bar\beta_j 
+ \bar\nabla_j\bar\beta_i\right),
\end{equation}
with $\bar\beta_i=\bar g_{ij}\beta^j$.  
The tracefree part of this equation implies
\begin{equation}
  \bar{A}^{ij}=\frac{1}{2\bar N}(\bar\Long\beta)^{ij}.
\end{equation}
Therefore, {\em with} the appropriate weight factor $\bar\sigma=2\bar
N$ as constructed above, the extrinsic curvature (\ref{eq:TT}) has
{\em no transverse traceless piece} for any spacelike slice in any
spacetime with a timelike Killing vector\footnote{A similar argument
is applicable in the ergosphere of a Kerr black hole; however, one
must be more careful with the choice of $\Sigma$ relative to $t$.}.  A
transverse-traceless decomposition of $\bar A^{ij}$ {\em without} the
weight-factor, however, will in general lead to a nonzero transverse
traceless piece.  This was previously a puzzle.  The TT part is
generally identified with the dynamical degrees of freedom.  Therefore
the radiative aspect of a stationary (or static) spacetime should be
manifestly zero on a natural slicing associated with the timelike
Killing vector.  The absence of this property for stationary,
non-static, spacetimes with previous decompositions was a serious
weakness.  

These considerations provide an independent argument for the
introduction of the weight-function in (\ref{eq:TT}), and the
identification of $\bar\sigma$ with the lapse-function in
(\ref{eq:sigma=2N}).

\section{A Geometrical Choice of \boldmath$N$ and \boldmath$\bar N$:
The conformal thin sandwich viewpoint}
\label{sec:VI}

The {\em conformal thin sandwich equations} \cite{York:1999} specify
freely (1) a conformal metric $g_{ij}$ and its velocity $\partial_t
g_{ij}=\dot g_{ij}$, and (2) the mean curvature $\trK$.  The lapse
$\bar N$ is $\CF^6 N$ with $N$ still adrift.  But there is a
definitive solution for fixing $N$: The mean curvature has become a
configuration variable, for which a value and a velocity need to be
specified; as one must specify $g_{ij}$ {\em and} its velocity $\dot
g_{ij}$, by analogy one can give the mean curvature $\trK$ {\em and}
its velocity $\partial_t\trK=\dot\trK$.  This will determine both $N$
and $\bar N$.

The specification 
\begin{equation}\label{eq:choice-new}
(g_{ij}, \dot g_{ij}; \trK,\dot\trK)
\end{equation}
 has the same {\em number} of variables as the conventional
choice 
\begin{equation}\label{eq:choice-old}
(\bar g_{ij}, \dot{\bar g}_{ij})=(g_{ij}, \dot g_{ij};
\bar g^{1/2}, \dot{\bar g}^{1/2}).
\end{equation}
  Furthermore $\bar g^{1/2}$ and $\trK$ are {\em canonically
conjugate} (apart from an irrelevant constant), so that
(\ref{eq:choice-new}) and (\ref{eq:choice-old}) are as close and as
symmetric to each other as possible.  However, the conventional
specification (\ref{eq:choice-old}) fails \cite{Bartnik-Fodor:1993}
while the conformal one (\ref{eq:choice-new}) does not fail.

Thus, in the conformal thin sandwich problem, we also give
\begin{align}
f[\beta; t, x]=\hat\partial_0\trK&=\partial_t\trK-\beta^i\partial_i\trK\\
\label{eq:d0-tau}
&=-\bar\Delta\bar N+(\bar R+\trK^2)\bar N
\end{align}
which is an Einstein equation on $\M$, conventionally regarded as an
equation for $\partial^2_t\bar g^{1/2}$. (Note that $\partial_t\bar
g^{1/2}\sim\trK$ is not a constraint; it is an identity, part of the
definition of extrinsic curvature [cf. (\ref{eq:dtgij})] ).  We are
turning the equation of motion for $\bar g^{1/2}$ into a constraint.
This is an old ``trick;'' $\trK=\dot\trK=0$ is maximal
slicing\cite{Lichnerowicz:1944}, whereas $\trK$ and $\dot\trK$
constant in space (but allowing changes in time) is constant mean
curvature slicing\cite{York:1972}.  Also, $\dot\trK=0$ is used during
construction of quasi-equilibrium initial data (see
e.g. \cite{Wilson-Matthews:1989,
Gourgoulhon-Grandclement-Bonazzola:2002a,Cook:2002} and references in
\cite{Cook:2000}).  However, now it is clear that specification of
$\dot\trK$ is fundamentally linked to the initial value problem.

We saw in the previous section that the choice of slicing ($\bar N$,
the passing of time) enters into the {\em construction} of the $(\bar
g_{ij}, \bar K_{ij})$ representation.  In addition, this
``Hamiltonian'' representation is consistent with the ``Lagrangian''
conformal thin sandwich picture.  Hence we can adopt (\ref{eq:d0-tau})
for the $(\bar g_{ij}, \bar K_{ij})$ representation as well.  In other
representations of the extrinsic curvature in this problem (Methods A
and B), there is no $\bar N$; and the construction has nothing to say
about the passage of time.  One does not move forward in time without
an extra equation to give $\bar N$.  Recall
$d\tau^{\mbox{\footnotesize (prop)}}/dt=\bar N$, where
$\tau^{\mbox{\footnotesize (prop)}}$ is a local Cauchy observer's
proper time and $t$ is the coordinate time.

Using the scalar constraint (\ref{eq:scalar-constraint0}) in
(\ref{eq:d0-tau}) yields (see e.g. Eqn. (98) of \cite{York:1979})
\begin{equation}
\begin{aligned}
f[\beta; t, x]
&=-\bar\Delta \bar N+\bar K_{ij}\bar K^{ij}\bar N\\
&=-\bar\Delta\bar N+\left(\bar A_{ij}\bar A^{ij}
+\mbox{$\frac{1}{3}$}\trK^2\right)\bar N.
\end{aligned}
\end{equation}
Conformal transformation of the Laplacian is carried out by expressing it as
\begin{equation}
\bar\Delta(\,.\,)
=\bar g^{-1/2}\partial_i\left[\bar g^{1/2}\bar g^{ij}\partial_j(\,.\,)\right].
\end{equation}
Preliminarily, we find
\begin{equation}
\begin{aligned}
f[\beta; t, x]
=-\CF^{-4}&\Delta(N\CF^6)-2\CF^{-5}(\nabla_i\CF)\left[\nabla^i(N\CF^6)\right]\\
 &+\Big(A_{ij}A^{ij}\,\CF^{-12}+\mbox{$\frac{1}{3}$}\trK^2\Big)(N\CF^6).
\end{aligned}
\end{equation}
To proceed, we use the scalar constraint in conformal form,
(\ref{eq:scalar-constraint3}),  to eliminate $\Delta\CF$ and find
\begin{equation}\label{eq:N}
\begin{aligned}
\Delta N&-\bigg[\frac{7}{4}A_{ij}A^{ij}\CF^{-8}-\frac{1}{6}\trK^2\CF^4-\frac{3}{4}R
  -42(\nabla_i\log\CF)^2\bigg]N\\
&+14(\nabla_iN)(\nabla^i\log\CF)
+\CF^{-2}(\partial_t\trK-\beta^i\partial_i\trK)=0,
\end{aligned}
\end{equation}
where $\trK(t,x)$ is given.\\

Equation (\ref{eq:N}) is a fifth elliptic equation, coupled to the
others, that is required for the completeness of the conformal thin
sandwich equations and is also natural in the extrinsic curvature
representation given here.

(Though the four conventional thin sandwich equations do not work, it
is interesting that in the Baierlein-Sharp-Wheeler (BSW) treatment
\cite{Baierlein-Sharp-Wheeler:1962}, there is an implicit fifth
equation.  Differentiation of the second order equation for the shift,
with a given lapse, produces a first order, {\em not} a third order,
equation: an integrability condition. This was discovered by Pereira
\cite{Pereira:1973}.)

\section{Conclusion}

By a simple tensor decomposition, one can bring the constraint
equations in the extrinsic curvature form into geometrical and
mathematical conformity with the conformal thin sandwich equations
when $N$ is arbitrary in both sets of equations.  This statement
remains true if $N$ is fixed by the same method in both formulations.

The conformal thin sandwich equations with $\dot\trK$ fixed form an
elliptic system whose general properties remain to be studied.  We
expect that it is a solvable, solid, system provided that $\trK$ and
$\dot\trK$ do not vary wildly.

A natural choice of $\bar N$ in stationary spacetimes has been seen to
render the transverse traceless part of $\bar K^{ij}$ zero.  The most
striking result of the present analysis is the inextricable relation
of time, represented by $\bar N$ or $\dot\trK$, and space, represented
by the four constraint equations.

\acknowledgments We thank Lawrence Kidder and Saul Teukolsky for
helpful discussions.  H. P. gives special thanks to Gregory Cook for
innumerable valuable discussions which shaped his understanding of the
initial value problem.  We gratefully acknowledge support by the
National Science Foundation through grants PHY-9800737, PHY-9900672,
PHY-9972582 and PHY-0216986.

\appendix*

\section{Second Fundamental Form and Extrinsic Curvature}
\label{app:second_form}

Let $\M$ be an $m$-dimensional surface embedded in a $d$-dimensional
ambient space $\V$ (We do not assume that $d=m+1$).  Let $\V$ be
endowed with a Riemannian or Lorentzian metric $\gfour$ and
corresponding Levi-Civita connection $D$, while $\M$ inherits a
Riemannian metric $g$ and connection $\nabla$.

Let $X$ and $Y$ be vectors in $\V$ that are tangent to
$\M$.  The first fundamental form of $\M$ for $X$ and
$Y$ is $g(X,Y)$, while the second fundamental form of $\M$
with respect to $X$ and $Y$ is the vector \cite{Taylor:1996}
\begin{equation}\label{eq:h}
  h(X,Y)\equiv \nabla_XY-D_XY.
\end{equation}
The {\em purpose of
the second fundamental form} is to discriminate between parallel
transport of a vector $Y$ along the direction of a vector $X$ in the
$(\V, \gfour, D)$ connection and in the $(\M, g,
\nabla)$ connection, when both $X$ and $Y$ are tangent to $\mathcal
M$.  This is defined without reference to surfaces near $\M$
or a foliation.  It tells us from the viewpoint of $\V$,
whether, say, the geodesics of $\M$ also appear ``straight''
in $\V$.  

Zero torsion in the Levi-Civita connections $D$ and $\nabla$ implies
\begin{equation}\label{eq:h-symmetric}
 h(X,Y)=h(Y,X).
\end{equation}

Now we demonstrate that $h(X,Y)$ is always orthogonal to $\M$.
Suppose $X, Y,$ and $Z$ are tangent to $\M$ and consider $\V$'s scalar
product between vectors $<\;,\;>=\gfour(\;,\;)$.  The product
rule for derivatives using $D_X$ gives
\begin{equation}\label{eq:productrule-h}
  <\!D_XY,Z\!>\,=X\!<\!Y,Z\!>-<\!Y,D_XZ\!>.
\end{equation}
A similar rearrangement using $\nabla_X$ gives the same expression
with $D_X$ replaced by $\nabla_X$.  Combining the two expressions so
as to cancel the common term $X<Y,Z>$, and invoking~(\ref{eq:h})
and~(\ref{eq:h-symmetric}) yields
\begin{equation}
<\!h(X,Y),Z\!>\,=-\!<\!h(Z,X),Y\!>\!.
\end{equation}
Hence, the trilinear form on the left changes sign under a cyclic
permutation of $X, Y,$ and $Z$.  But three such permutations restore
the original order, which must then be the negative of itself.  Hence,
it is zero, and thus $h(X,Y)$ is orthogonal to $\M$.

Let us state the consequences in tensor language by supposing that
$\M$ has an adapted basis $e_i$ ($i,j,\ldots=1,2,\ldots,m$).  The full
basis of $\V$ is $e_\alpha$ ($\alpha,\beta,\ldots=1,2,\ldots,d$),
where the first $m$ vectors are the $e_i$.

We define the coefficients of the connection one-forms of $(\V, \gfour,
D)$ by
\begin{equation}
D_{e_\alpha}e_\beta=\omega_{\alpha\beta}^\gamma e_\gamma
\end{equation}
(this differs from the convention of Misner, Thorn, and Wheeler (MTW)
\cite{Misner-Thorne-Wheeler:1973}).  After a brief calculation we find
that $h(X, Y)$ can be written as
\begin{equation}\label{eq:h-components}
  h(X,Y)
=-\sum_{\alpha=m+1}^d e_\alpha \omega^\alpha_{ij}X^iY^j,
\end{equation}
We expand the basis of the co-space of $\M$ in $\V$ in terms of
$(d-m)$ mutually orthogonal unit normals to $\M$, $n_{\hat a}$, $\hat
a=\hat m+1, \ldots,\hat d$.  There is a $(d-m)\times(d-m)$ nonsingular
matrix $E^{\hat a}_\alpha$ such that
\begin{equation}
e_\alpha=\sum_{\hat a=\hat m+\hat 1}^{\hat d}E^{\hat a}_\alpha n_{\hat a},\qquad \alpha=m+1,\ldots,d.
\end{equation}
Hence, (\ref{eq:h-components}) becomes
\begin{equation}
h(X,Y)
=-\sum_{\alpha=m+1}^d\;\sum_{\hat a=\hat m+1}^{\hat d}
    n_{\hat a}\left(E^{\hat a}_\alpha\omega^\alpha_{ij}\right)X^iY^j.
\end{equation}

The $d-m$ extrinsic curvature tensors $K^{\hat a}_{\,ij}$ are defined by
\begin{equation}\label{eq:def-Kij}
  K^{\hat a}_{\,ij}=-\sum_{\alpha=m+1}^dE^{\hat a}_{\alpha}\omega^\alpha_{ij}.
\end{equation}

Equations (\ref{eq:def-Kij}) and (\ref{eq:h}) emphasize that the
extrinsic curvature is related to transport {\em parallel} to
the slice, a viewpoint not present in a definition in terms of
derivatives {\em normal} to the slice.  But both definitions
agree.
\\

Now let $\gfour$ be Lorentzian, $g$ Riemannian, and $m=d-1$.  We are
interested in the case $d=4, m=3$. In this case $\M$ is a hypersurface,
which we take to be $t=\mbox{const}$.  There is only one extrinsic
curvature tensor $K_{ij}$.  The spacetime metric is
\begin{equation}\label{eq:metric}
\gfour=-N^2dt^2+g_{ij}(dx^i+\beta^idt)(dx^j+\beta^jdt),
\end{equation}
where $N$ is the lapse and $\beta^i$ the shift.  We choose the
``Cauchy-adapted'' coframe $\theta^\alpha$ ($\alpha, \beta, \ldots
=0,1,2,3$; $i,j,\ldots=1,2,3$):
\begin{equation}\label{eq:coframe}
  \theta^0=dt,\qquad\qquad \theta^i=dx^i+\beta^idt.
\end{equation}
The dual vector frame is $e_\alpha=\partial_\alpha$, with
\begin{equation}\label{eq:dual-frame}
  \partial_0=\partial_t-\beta^i\partial_i,\qquad\qquad 
  \partial_i=\dnachd{}{x^i}.
\end{equation}
We use $\partial$ to denote Pfaffian derivatives, some of which are
natural (namely $\partial_i=\partial/\partial x^i$ and
$\partial_t=\partial/\partial t$). In particular, the spatial basis
$\partial_i$ is natural, so that the connection coefficients and
Christoffel-symbols of $g_{ij}$ are equal.

 For the hypersurface $t=\mbox{const.}$ we find
\begin{align}
  \nabla_{X}Y-D_{X}Y 
&= -X^iY^j(\omega^0_{ij}e_0)\nonumber\\
&= -X^iY^j(N\omega^0_{ij})n_{\hat 0},
\end{align}
where we used $e_0=Nn_{\hat 0}$.  $\omega_{ij}^0$ can be
evaluated by the formula
\begin{align}
\omega_{\beta\gamma}^\alpha
&=\Gamma^\alpha_{\beta\gamma}
  +\frac{1}{2}\gfour^{\alpha\delta}\left(C^\lambda_{\delta\beta}
       \gfour_{\gamma\lambda}
       +C^\lambda_{\delta\gamma}
       \gfour_{\beta\lambda}\right)
  +\frac{1}{2}C^\alpha_{\beta\gamma},
\end{align}
where the structure coefficients $C^\alpha_{\beta\gamma}$ are defined by
\begin{equation}
d\theta^\alpha
=-\frac{1}{2}C^\alpha_{\beta\gamma}\,\theta^\beta\wedge\theta^\gamma;
\qquad
[e_\alpha, e_\beta]=C^\gamma_{\alpha\beta}e_\gamma.
\end{equation}
In our frame, all $C^\alpha_{\beta\gamma}$ vanish except
$C^i_{0j}=-C^i_{j0}=\partial_j\beta^i$, and one finds
\begin{equation}\label{eq:K0ij}
  K_{ij}\equiv K^{\hat 0}_{\;ij}
=-N\omega^0_{ij}
=-\frac{1}{2}N^{-1}\left(\partial_t-\Lie_\beta\right)g_{ij}.
\end{equation}
Here, $\Lie_\beta$ is the spatial Lie derivative along the shift
vector.  Since $\Lie_\beta g_{ij}=\nabla_i\beta_j+\nabla_j\beta_i$
with $\beta_i\equiv g_{ij}\beta^j$, (\ref{eq:K0ij}) gives
\begin{equation}\label{eq:Kij}
  K_{ij}=-\frac{1}{2}N^{-1}\big(\partial_tg_{ij}
         -\nabla_i\beta_j-\nabla_j\beta_i\big).
\vspace*{0.4em}
\end{equation}
Rewriting (\ref{eq:Kij}) gives
\begin{equation}\label{eq:dtgij}
  \partial_tg_{ij}=-2NK_{ij}+\nabla_i\beta_j+\nabla_j\beta_i.
\end{equation}
If you prefer the opposite sign for $K_{ij}$, as some authors do,
simply change the sign of $h(X,Y)$ in its definition.  Equations
(\ref{eq:Kij}) and (\ref{eq:dtgij}) change sign when passing from the
Lorentzian to the Riemannian (``Euclidean'') case for either choice of
the sign of $h(X,Y)$.  
\\

For completeness, we give all connection coefficients of the
frame defined by (\ref{eq:metric}), (\ref{eq:coframe}) and
(\ref{eq:dual-frame}): 
\begin{align}
\omega^0_{00}&=\partial_0\log N,&\omega^0_{ij}&=-N^{-1}K_{ij},\\
\label{eq:A20}
\omega^i_{j0}&=-NK^i_{\,j},&\omega^i_{0j}&=-NK^i_{\,j}+\partial_j\beta^i,\\
\omega^0_{i0}&=\omega^0_{0i}=\partial_i\log N,
& \omega^i_{00}&=N g^{ij}\partial_jN,\\
\omega^i_{jk}&=\Gamma^i_{jk}.
\end{align}
We have rewritten time-derivatives in terms of the extrinsic
curvature; $\Gamma^i_{jk}$ denotes the Christoffel-symbols of the
spatial metric $g_{ij}$.  We note again that we do not use the MTW
convention for the order of indices of the connection coefficients.
In our frame, this is significant only for (\ref{eq:A20}).

\bibliography{york}

\begin{thebibliography}{29}
\expandafter\ifx\csname natexlab\endcsname\relax\def\natexlab#1{#1}\fi
\expandafter\ifx\csname bibnamefont\endcsname\relax
  \def\bibnamefont#1{#1}\fi
\expandafter\ifx\csname bibfnamefont\endcsname\relax
  \def\bibfnamefont#1{#1}\fi
\expandafter\ifx\csname citenamefont\endcsname\relax
  \def\citenamefont#1{#1}\fi
\expandafter\ifx\csname url\endcsname\relax
  \def\url#1{\texttt{#1}}\fi
\expandafter\ifx\csname urlprefix\endcsname\relax\def\urlprefix{URL }\fi
\providecommand{\bibinfo}[2]{#2}
\providecommand{\eprint}[2][]{\url{#2}}

\bibitem[{\citenamefont{York{,}~Jr.}(1979)}]{York:1979}
\bibinfo{author}{\bibfnamefont{J.~W.} \bibnamefont{York{,}~Jr.}}, in
  \emph{\bibinfo{booktitle}{Sources of Gravitational Radiation}}, edited by
  \bibinfo{editor}{\bibfnamefont{L.~L.} \bibnamefont{Smarr}}
  (\bibinfo{address}{Cambridge University Press, Cambridge, England},
  \bibinfo{year}{1979}), p.~\bibinfo{pages}{83}.

\bibitem[{\citenamefont{Murchadha and
  York{,}~Jr.}(1974)}]{Murchadha-York:1974a}
\bibinfo{author}{\bibfnamefont{N.~O.} \bibnamefont{Murchadha}}
  \bibnamefont{and} \bibinfo{author}{\bibfnamefont{J.~W.}
  \bibnamefont{York{,}~Jr.}}, \bibinfo{journal}{Phys. Rev. D}
  \textbf{\bibinfo{volume}{10}}, \bibinfo{pages}{428} (\bibinfo{year}{1974}).

\bibitem[{\citenamefont{York{,}~Jr.}(1999)}]{York:1999}
\bibinfo{author}{\bibfnamefont{J.~W.} \bibnamefont{York{,}~Jr.}},
  \bibinfo{journal}{Phys. Rev. Lett.} \textbf{\bibinfo{volume}{82}},
  \bibinfo{pages}{1350} (\bibinfo{year}{1999}).

\bibitem[{\citenamefont{Deser}(1967)}]{Deser:1967}
\bibinfo{author}{\bibfnamefont{S.}~\bibnamefont{Deser}}, \bibinfo{journal}{Ann.
  Inst. Henri Poincar{\'e}, Section A} \textbf{\bibinfo{volume}{7}},
  \bibinfo{pages}{149} (\bibinfo{year}{1967}).

\bibitem[{\citenamefont{York{, Jr.}}(1973)}]{York:1973}
\bibinfo{author}{\bibfnamefont{J.~W.} \bibnamefont{York{, Jr.}}},
  \bibinfo{journal}{J. Math. Phys.} \textbf{\bibinfo{volume}{14}},
  \bibinfo{pages}{456} (\bibinfo{year}{1973}).

\bibitem[{\citenamefont{York{, Jr.}}(1974)}]{York:1974}
\bibinfo{author}{\bibfnamefont{J.~W.} \bibnamefont{York{, Jr.}}},
  \bibinfo{journal}{Ann. Inst. Henri Poincar{\'e}, Section A}
  \textbf{\bibinfo{volume}{21}}, \bibinfo{pages}{319} (\bibinfo{year}{1974}).

\bibitem[{\citenamefont{Pfeiffer et~al.}(2002)\citenamefont{Pfeiffer, Cook, and
  Teukolsky}}]{Pfeiffer-Cook-Teukolsky:2002}
\bibinfo{author}{\bibfnamefont{H.~P.} \bibnamefont{Pfeiffer}},
  \bibinfo{author}{\bibfnamefont{G.~B.} \bibnamefont{Cook}}, \bibnamefont{and}
  \bibinfo{author}{\bibfnamefont{S.~A.} \bibnamefont{Teukolsky}},
  \bibinfo{journal}{Phys. Rev. D} \textbf{\bibinfo{volume}{66}},
  \bibinfo{pages}{024047} (\bibinfo{year}{2002}).

\bibitem[{\citenamefont{Murchadha}(1973)}]{Murchadha:1973}
\bibinfo{author}{\bibfnamefont{N.~{\'O}.} \bibnamefont{Murchadha}}
  (\bibinfo{year}{1973}), \bibinfo{note}{private communication}.

\bibitem[{\citenamefont{Isenberg}(1974)}]{Isenberg:1974}
\bibinfo{author}{\bibfnamefont{J.}~\bibnamefont{Isenberg}}
  (\bibinfo{year}{1974}), \bibinfo{note}{private communication}.

\bibitem[{\citenamefont{Cantor}(1980)}]{Cantor:1980}
\bibinfo{author}{\bibfnamefont{M.}~\bibnamefont{Cantor}}
  (\bibinfo{year}{1980}), \bibinfo{note}{private communication}.

\bibitem[{\citenamefont{Cook}(2000)}]{Cook:2000}
\bibinfo{author}{\bibfnamefont{G.~B.} \bibnamefont{Cook}},
  \bibinfo{journal}{Living Rev. Relativity} \textbf{\bibinfo{volume}{3}}
  (\bibinfo{year}{2000}), \bibinfo{note}{[Online Article: cited on Jul 15,
  2002]},
  \urlprefix\url{http://livingreviews.org/Articles/Volume3/2000-5cook/}.

\bibitem[{\citenamefont{Lichnerowicz}(1944)}]{Lichnerowicz:1944}
\bibinfo{author}{\bibfnamefont{A.}~\bibnamefont{Lichnerowicz}},
  \bibinfo{journal}{J. Math. Pures \& Appl.} \textbf{\bibinfo{volume}{23}},
  \bibinfo{pages}{37} (\bibinfo{year}{1944}).

\bibitem[{\citenamefont{Arnowitt et~al.}(1962)\citenamefont{Arnowitt, Deser,
  and Misner}}]{Arnowitt-Deser-Misner:1962}
\bibinfo{author}{\bibfnamefont{R.}~\bibnamefont{Arnowitt}},
  \bibinfo{author}{\bibfnamefont{S.}~\bibnamefont{Deser}}, \bibnamefont{and}
  \bibinfo{author}{\bibfnamefont{C.~W.} \bibnamefont{Misner}}, in
  \emph{\bibinfo{booktitle}{Gravitation: An introduction to Current Research}},
  edited by \bibinfo{editor}{\bibfnamefont{L.}~\bibnamefont{Witten}}
  (\bibinfo{publisher}{Wiley, New York}, \bibinfo{year}{1962}), p.
  \bibinfo{pages}{227}.

\bibitem[{\citenamefont{Choquet-Bruhat and
  Ruggeri}(1983)}]{Choquet-Bruhat-Ruggeri:1983}
\bibinfo{author}{\bibfnamefont{Y.}~\bibnamefont{Choquet-Bruhat}}
  \bibnamefont{and} \bibinfo{author}{\bibfnamefont{T.}~\bibnamefont{Ruggeri}},
  \bibinfo{journal}{Commun. Math. Phys.} \textbf{\bibinfo{volume}{89}},
  \bibinfo{pages}{269} (\bibinfo{year}{1983}).

\bibitem[{\citenamefont{Teitelboim}(1983)}]{Teitelboim:1983}
\bibinfo{author}{\bibfnamefont{C.}~\bibnamefont{Teitelboim}},
  \bibinfo{journal}{Phys. Rev. D} \textbf{\bibinfo{volume}{28}},
  \bibinfo{pages}{297} (\bibinfo{year}{1983}).

\bibitem[{\citenamefont{Ashtekar}(1987)}]{Ashtekar:1987}
\bibinfo{author}{\bibfnamefont{A.}~\bibnamefont{Ashtekar}},
  \bibinfo{journal}{Phys. Rev. D} \textbf{\bibinfo{volume}{36}},
  \bibinfo{pages}{1587} (\bibinfo{year}{1987}).

\bibitem[{\citenamefont{Anderson and York{, Jr.}}(1998)}]{Anderson-York:1998}
\bibinfo{author}{\bibfnamefont{A.}~\bibnamefont{Anderson}} \bibnamefont{and}
  \bibinfo{author}{\bibfnamefont{J.~W.} \bibnamefont{York{, Jr.}}},
  \bibinfo{journal}{Phys. Rev. Lett.} \textbf{\bibinfo{volume}{81}},
  \bibinfo{pages}{1154} (\bibinfo{year}{1998}).

\bibitem[{\citenamefont{York{, Jr.}}(1999)}]{York:1999b}
\bibinfo{author}{\bibfnamefont{J.~W.} \bibnamefont{York{, Jr.}}}, in
  \emph{\bibinfo{booktitle}{Relativity, Particle Physics and Cosmology}},
  edited by \bibinfo{editor}{\bibfnamefont{R.~E.} \bibnamefont{Allen}}
  (\bibinfo{publisher}{World Sientific, Singapore}, \bibinfo{year}{1999}), pp.
  \bibinfo{pages}{77--85}.

\bibitem[{\citenamefont{Choquet-Bruhat and
  York}(1997)}]{Choquet-Bruhat-York:1997}
\bibinfo{author}{\bibfnamefont{Y.}~\bibnamefont{Choquet-Bruhat}}
  \bibnamefont{and} \bibinfo{author}{\bibfnamefont{J.~W.} \bibnamefont{York}},
  \bibinfo{journal}{Banach Center Publ.} \textbf{\bibinfo{volume}{41}},
  \bibinfo{pages}{119} (\bibinfo{year}{1997}).

\bibitem[{\citenamefont{Kidder et~al.}(2001)\citenamefont{Kidder, Scheel, and
  Teukolsky}}]{Kidder-Scheel-Teukolsky:2001}
\bibinfo{author}{\bibfnamefont{L.~E.} \bibnamefont{Kidder}},
  \bibinfo{author}{\bibfnamefont{M.~A.} \bibnamefont{Scheel}},
  \bibnamefont{and} \bibinfo{author}{\bibfnamefont{S.~A.}
  \bibnamefont{Teukolsky}}, \bibinfo{journal}{Phys. Rev. D}
  \textbf{\bibinfo{volume}{64}}, \bibinfo{pages}{064017}
  (\bibinfo{year}{2001}).

\bibitem[{\citenamefont{Bartnik and Fodor}(1993)}]{Bartnik-Fodor:1993}
\bibinfo{author}{\bibfnamefont{R.}~\bibnamefont{Bartnik}} \bibnamefont{and}
  \bibinfo{author}{\bibfnamefont{G.}~\bibnamefont{Fodor}},
  \bibinfo{journal}{Phys. Rev. D} \textbf{\bibinfo{volume}{48}},
  \bibinfo{pages}{3596} (\bibinfo{year}{1993}).

\bibitem[{\citenamefont{York{, Jr.}}(1972)}]{York:1972}
\bibinfo{author}{\bibfnamefont{J.~W.} \bibnamefont{York{, Jr.}}},
  \bibinfo{journal}{Phys. Rev. Lett.} \textbf{\bibinfo{volume}{28}},
  \bibinfo{pages}{1082} (\bibinfo{year}{1972}).

\bibitem[{\citenamefont{Wilson and Mathews}(1989)}]{Wilson-Matthews:1989}
\bibinfo{author}{\bibfnamefont{J.~R.} \bibnamefont{Wilson}} \bibnamefont{and}
  \bibinfo{author}{\bibfnamefont{G.~J.} \bibnamefont{Mathews}}, in
  \emph{\bibinfo{booktitle}{Frontiers in Numerical Relativity}}, edited by
  \bibinfo{editor}{\bibfnamefont{C.~R.} \bibnamefont{Evans}},
  \bibinfo{editor}{\bibfnamefont{L.~S.} \bibnamefont{Finn}}, \bibnamefont{and}
  \bibinfo{editor}{\bibfnamefont{D.~W.} \bibnamefont{Hobill}}
  (\bibinfo{publisher}{Cambridge University Press},
  \bibinfo{address}{Cambridge, England, 1989}, \bibinfo{year}{1989}), pp.
  \bibinfo{pages}{306--314}.

\bibitem[{\citenamefont{Gourgoulhon et~al.}(2002)\citenamefont{Gourgoulhon,
  Grandcl{\'e}ment, and Bonazzola}}]{Gourgoulhon-Grandclement-Bonazzola:2002a}
\bibinfo{author}{\bibfnamefont{E.}~\bibnamefont{Gourgoulhon}},
  \bibinfo{author}{\bibfnamefont{P.}~\bibnamefont{Grandcl{\'e}ment}},
  \bibnamefont{and}
  \bibinfo{author}{\bibfnamefont{S.}~\bibnamefont{Bonazzola}},
  \bibinfo{journal}{Phys. Rev. D} \textbf{\bibinfo{volume}{65}},
  \bibinfo{pages}{044020} (\bibinfo{year}{2002}).

\bibitem[{\citenamefont{Cook}(2002)}]{Cook:2002}
\bibinfo{author}{\bibfnamefont{G.~B.} \bibnamefont{Cook}},
  \bibinfo{journal}{Phys. Rev. D} \textbf{\bibinfo{volume}{65}},
  \bibinfo{pages}{084003} (\bibinfo{year}{2002}).

\bibitem[{\citenamefont{Baierlein et~al.}(1962)\citenamefont{Baierlein, Sharp,
  and Wheeler}}]{Baierlein-Sharp-Wheeler:1962}
\bibinfo{author}{\bibfnamefont{R.~F.} \bibnamefont{Baierlein}},
  \bibinfo{author}{\bibfnamefont{D.~H.} \bibnamefont{Sharp}}, \bibnamefont{and}
  \bibinfo{author}{\bibfnamefont{J.~A.} \bibnamefont{Wheeler}},
  \bibinfo{journal}{Phys. Rev.} \textbf{\bibinfo{volume}{126}},
  \bibinfo{pages}{1864} (\bibinfo{year}{1962}).

\bibitem[{\citenamefont{Pereira}(1973)}]{Pereira:1973}
\bibinfo{author}{\bibfnamefont{C.}~\bibnamefont{Pereira}}, \bibinfo{journal}{J.
  Math. Phys.} \textbf{\bibinfo{volume}{14}}, \bibinfo{pages}{1498}
  (\bibinfo{year}{1973}).

\bibitem[{\citenamefont{Taylor}(1996)}]{Taylor:1996}
\bibinfo{author}{\bibfnamefont{M.~E.} \bibnamefont{Taylor}},
  \emph{\bibinfo{title}{Partial Differential Equations II}}
  (\bibinfo{publisher}{Springer}, \bibinfo{address}{Berlin},
  \bibinfo{year}{1996}), \bibinfo{note}{p. 479}.

\bibitem[{\citenamefont{Misner et~al.}(1973)\citenamefont{Misner, Thorne, and
  Wheeler}}]{Misner-Thorne-Wheeler:1973}
\bibinfo{author}{\bibfnamefont{C.~W.} \bibnamefont{Misner}},
  \bibinfo{author}{\bibfnamefont{K.~S.} \bibnamefont{Thorne}},
  \bibnamefont{and} \bibinfo{author}{\bibfnamefont{J.~A.}
  \bibnamefont{Wheeler}}, \emph{\bibinfo{title}{Gravitation}}
  (\bibinfo{publisher}{W. H. Freeman, New York}, \bibinfo{year}{1973}).

\end{thebibliography}

\end{document}